\documentclass
[prl,superscriptaddress,twocolumn,bibnotes,notitlepage,preprint,10pt]{revtex4}%
\usepackage{amsfonts}
\usepackage{amsmath}
\usepackage{amssymb}
\usepackage{graphicx}%
\begin{document}
\preprint{ }

\textbf{Comment on \textquotedblleft Nanoconfinement-Enhanced Conformational
Response of Single DNA Molecules to Changes in Ionic
Environment\textquotedblright}

In a recent Letter \cite{Reisner}, Reisner \textit{et al.} study the ionic
strength dependence of DNA extension in nanochannels.\ One of the central
claims of the work is that the variation of DNA persistence length with ionic
strength is alone insufficient to explain the observed variation of DNA
extension. To this end, the authors argue that it is necessary to introduce an
ionic-strength dependent effective polymer width to account for the effects of
increased self-avoidance. This is plausible at high ionic strengths and large
channel cross-sections, i.e., when the DNA polymer is expected to exhibit
blob-like behavior \cite{deGennes}. However, in the regime of low ionic
strength, where the interaction of the DNA backbone with the confining channel
walls precludes intersegmental repulsion, the relevance of an effective
polymer width is debatable and warrants some scrutiny. In the present Comment,
we argue that the data of Reisner \textit{et al.} at low ionic strengths
$I\lesssim30$ mM are in fact described very well without the notion of an
effective polymer width using the entropic depletion theory due to Odijk
\cite{Odijk2006} with an ionic-strength dependent DNA persistence length.

\begin{figure}
[ptb]
\begin{center}
\includegraphics[
height=1.3136in,
width=3.3183in
]%
{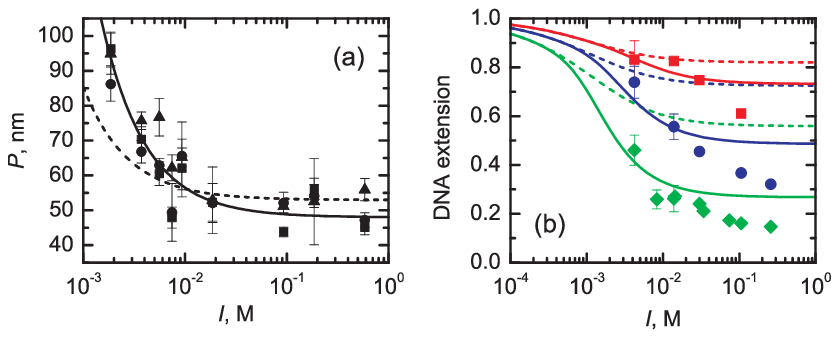}%
\caption{(color online) (a) Experimental data of \cite{Baumann1997} (symbols)
and their weighted least-squares fit to an OSF-type formula yielding the
dependence $P=\left(  48.0+\left(  0.0849\text{M}/I\right)  \right)  $ nm
($-$); also shown is the fit to the standard OSF formula \cite{Odijk1977}
\cite{SkolnickFixman} $P=\left(  P_{0}+\left(  0.0324\text{M}/I\right)
\right)  $ nm with $P_{0}$ as a free parameter\ (- - -). See also discussion
in\ \cite{Manning2006}. (b) Experimental data for DNA extension in
nanochannels from \cite{Reisner} (nanochannel cross-section side length
$\approx$50 nm ($\blacksquare$), $\approx$100 nm ($\bullet$), $\approx$200 nm
($\blacklozenge$)) superimposed with the predictions of the theory
\cite{Odijk2006} supplemented with the ionic-strength dependence of the DNA
persistence length derived from data of \cite{Baumann1997} ($-$); for
comparison, the corresponding predictions of the theory \cite{Odijk1983} are
shown (- - -).}%
\label{Fig. 1}%
\end{center}
\end{figure}

The conclusion of Reisner \textit{et al. }that the model based solely on
persistence length fails to explain the experimental observations stems from a
comparison of the DNA extension data with their Eq. (4) derived in
\cite{Odijk1983}. However, this expression is expected to hold only when the
extended length $r$ of DNA confined in a nanochannel approaches its contour
length $L$, i.e. $\left(  L-r\right)  /L\ll1$. It is apparent from Fig. 1 of
\cite{Reisner} that this condition is not satisfied for a majority of the data
points. Therefore, we compare these data with the expression derived in
\cite{Odijk2006} (incidentally, also cited in \cite{Reisner}), which is free
from the above approximation. In this case, the extended length of DNA in a
square nanochannel with the cross-section of side $A$ is $r=R_{e}\left\langle
\cos\eta\right\rangle =R_{e}\left(  1-\left\langle \eta^{2}\right\rangle
/2\right)  ,$ where $\left\langle \eta^{2}\right\rangle =0.340\left(
A/P\right)  ^{2/3}$, $R_{e}^{2}=2Lg-2g^{2}\left(  1-\exp\left(  -L/g\right)
\right)  $, $L$ is the polymer contour length, and $g$ is\ the global
persistence length (for details, see \cite{Odijk2006}). The experimental data
on the ionic strength dependence of the DNA persistence length relevant to
single-molecule DNA\ experiments obtained in \cite{Baumann1997} are described
well by an Odijk$-$Skolnick$-$Fixman (OSF) type formula \cite{Odijk1977}
\cite{SkolnickFixman} (see Fig. \ref{Fig. 1}a). Using this dependence, we
compare the theory \cite{Odijk2006} with the data of \cite{Reisner} and find
very good agreement at low ionic strengths (Fig. \ref{Fig. 1}b).%

%TCIMACRO{\FRAME{ftbpFU}{3.3183in}{1.3136in}{0pt}{\Qcb{(color online) (a)
%Experimental data of \cite{Baumann1997} (symbols) and their weighted
%least-squares fit to an OSF-type formula yielding the dependence $P=\left(
%48.0+\left(  0.0849\text{M}/I\right)  \right)  $ nm ($-$); also shown is the
%fit to the standard OSF formula \cite{Odijk1977} \cite{SkolnickFixman}
%$P=\left(  P_{0}+\left(  0.0324\text{M}/I\right)  \right)  $ nm with $P_{0}$
%as a free parameter\ (- - -). See also discussion in\ \cite{Manning2006}. (b)
%Experimental data for DNA extension in nanochannels from \cite{Reisner}
%(nanochannel cross-section side length $\approx$50 nm ($\blacksquare$),
%$\approx$100 nm ($\bullet$), $\approx$200 nm ($\blacklozenge$)) superimposed
%with the predictions of the theory \cite{Odijk2006} supplemented with the
%ionic-strength dependence of the DNA persistence length derived from data of
%\cite{Baumann1997} ($-$); for comparison, the corresponding predictions of the
%theory \cite{Odijk1983} are shown (- - -).}}{\Qlb{Fig. 1}}{Fig1.eps}%
%{\special{ language "Scientific Word";  type "GRAPHIC";
%maintain-aspect-ratio TRUE;  display "USEDEF";  valid_file "F";
%width 3.3183in;  height 1.3136in;  depth 0pt;  original-width 5.4717in;
%original-height 2.2407in;  cropleft "0";  croptop "1";  cropright "1";
%cropbottom "0";  filename 'Fig1.eps';file-properties "XNPEU";}}}%
%BeginExpansion
%EndExpansion

Furthermore, in order to describe their data, the authors of \cite{Reisner}
invoke the scaling-based de\ Gennes theory \cite{deGennes} which, as it turns
out, is quite insensitive to particular functional dependences of both the DNA
persistence length and width on the ionic strength. In fact, agreement with
data for $I>10^{-2}$ M can be achieved by selecting an appropriate numerical
prefactor \cite{Reisner}, although it still fails for lower $I$. By contrast,
Odijk's theory \cite{Odijk2006} is extremely sensitive to the particular
functional dependence $P\left(  I\right)  $ and has no tunable parameters. It
not only predicts the correct magnitude of the effect, but also describes
experimental data well for $I\lesssim30$ mM (Fig. \ref{Fig. 1}b). Note that
this theory is not expected to describe the data at higher ionic strengths.
Thus, we demonstrate that contrary to the conclusions made by Reisner
\textit{et al.} \cite{Reisner}, the ionic strength dependence of the DNA
persistence length alone is indeed sufficient to explain the experimental data
on DNA extension in nanochannels at low ionic strengths.

We thank Theo Odijk for illuminating discussions.

Madhavi Krishnan and Eugene P. Petrov

Biophysics, BIOTEC, Technische Universit\"{a}t Dresden, Tatzberg 47-51, 01307
Dresden, Germany

Received \today


\end{document}